# A Recent Survey of Event Triggered Control of Nonlinear Systems

Raffay Yaqoob

School of Automation, *Central south University, Changsha 10533, China*

**Abstract**

The control systems are an essential part of every engineering system in any industrial application. The basic purpose of controls is to manage the internal operations of the system and detect any unwanted or uncertain situation. Failure in any system may be caused by various reasons and need to be restricted to save other components of the system. event-triggered control systems manage errors and failures by identifying particular events caused by failure and then mitigate the effect of that failure as much as possible. These systems have become prominent due to their smooth and autonomous way of detecting and minimizing risk in any system. The input to state stability property and a perfectly stabilized control are assumed for applying event-triggered control systems. Switching topologies can be utilized as an alternative to ISS assumptions to design triggering events. There are some issues related to this approach like non- linearity of systems, multi-agent communication, heterogeneity and Zeno effect in systems while monitoring, controlling faults and taking feedbacks. The efficiency of neural networks and fuzzy technologies are considered in this regard which are investigated by many researchers in the last decade. This survey reviews all work related to event-triggered control systems, their applications, challenges and possible solutions. The adaptability of these controls is evaluated based on the solutions available as well as the applicability of solutions proposed by researchers.

**Keywords**

Event Triggered Control, neural network, nonlinear systems, fault control, feedback systems, multiagent systems

## 1. Introduction

Engineering controls are used in various applications such as machine controls, process monitoring, fabrication, event management, security, data processing etc. Failures are a part of systems, which can be triggered due to ageing, wear, electrical issues, sensor failures, actuator failures, internal uncertainties and various other reasons(M. Li & Chen, 2019; Ma & Ma, 2018). Adaptive technologies assist in controlling the adverse effects of failure by recognizing the cause and converting the behavior accordingly to accommodate the effects due to failure of a component(Triki-Lahiani, Bennani-Ben Abdelghani, &Slama-Belkhodja, 2018; Yoon, Boragule, Song, Yoon, & Jeon, 2018). Event trigger control systems tend to control the issues by recognizing the event and mitigating the issue caused by event using an appropriate strategy. The fault is studied and analyzed by the system automatically by implementing of adaptive neural networks or fuzzy technologies(G. Liu, Cao, & Chang, 2017; M. Liu, Yu, Sun, & Li, 2021; Ziyabari&Shoorehdeli, 2017).

With the immense development of networking systems and information technology, the many of the communication devices are not connected to the plants and control systems using networking channels. This trend has reached its peak during recent pandemic where internet became the basic mode of communication and operations of the systems. Network based controlled are used in smart grids, multi-agent systems, faraway platforms, security systems, defense and others. The device failure was mainly responsible for the network failure however, communication burden and costs to replace all the system or to down a communication completely for some time is usually undesired. Network controlling is usually closely associated with communication systems, control science, information technology, internet of things, etc. but their applications is limited due to limited network bandwidth.

Event triggering mechanism is a prospective approach that improves the implementation feasibility of network control systems. This system may be used in cyber physical system, multi-agent system, and distributed control systems. The results of event triggered controls can be assessment conveniently by assuming that the control is perfectly stable and the input to state stability property is present. Event based controller checked for various environmental triggers and reacts accordingly to manage the resources and workload effectively. The ISS and triggering events are also designed using switching topologies without using ISS assumptions. The use of high order non linear system controllers with both dynamic and static mechanisms for trigger of events are used in research to prove the global stability of event triggering mechanism and its effectiveness in network control.

Neural networks (NN) are used to assess the results of application of event trigger management system on networking controls tracking and regulation of various non linear systems. Researchers have proposed several trigger based control for uncertain nonlinear functions which are further extended for application in Adding nonlinear systems. Reinforcement learning and adaptive dynamic programming is added to reinforcement learning in the preparation of the controller.

Historically, the earlier developed ETC focused on unpredictable nonlinear systems with predictable matching conditions. This trend shifted to the research on ETC on unpredictable nonlinear systems with unpredictable matching conditions. A researcher developed control law for nonlinear dynamic model for control systems using state estimation and approximation set by neural network. Finally, the research shifted its focus towards the new neural networks and its weights employing the mismatch between system and model matrix.

This review paper comprises summary of previously published research along with the necessary details to serve as a bridge between the new research, section I describes the event triggered adaptive control and communication of uncertain nonlinear systems and section II describes the research on event triggered control of multiagent systems.

**Section I**

**Event-triggered Adaptive Control for a Class of Uncertain Nonlinear Systems**

In this research, the researcher has introduced the new switching threshold strategy by combining the constraints of fixed threshold strategy and relative threshold strategy. He has combined both strategies so that we can get the precise control system and to ensure certain system performances by keeping the bounded measurements. He has provided a protocol to design the adoptive controller and triggering mechanism at the same time to compensate for the measurement errors and ISS assumptions is not required. Also, he has given examples of tracking case simulation and stabilization case simulation in which all three strategies are applied and their results are compared with their effectiveness. Based on the examples, it clearly shows that the proposed switching threshold strategy is best for balancing the system constraints (L. Xing, C. Wen, Z. Liu, H. Su and J. Cai, 2017).

**Event-Triggered Adaptive Control of Uncertain Nonlinear Systems with Composite Condition**

In this research, the researcher has introduced the event triggered adoptive control with composite triggering condition and Neural Network weights composite adoptive law. By designing the controller based on NN weights adoptive law are updated only at triggering events. Event base adaptive control is discussed in details along with Event triggering condition and stability analysis. The triggering events are greatly reduced by applying this technique which in further reduce the triggering errors at reduced events. Lyapunov analysis is also applied to prove the stability of the system. Simulation examples are given at the end which clearly shows that data sampling instants are reduced(X. Liu, Xu, Shou, Fan, & Chen, 2021).

**Event-Triggered Global Finite-Time Control for a Class of Uncertain Nonlinear Systems**

In this paper, the researcher has discussed the global finite time stabilization problem and how to overcome this issue. Existing methods can only partially reduce the event triggered errors. A new technique for event triggering mechanism and controller co-design based on back-stepping and sign function technique. This technique is also proved by applying the Lyapunov analysis and finite time stability theory which gives us the finite time stability of the system under observation. Examples are given at the end with calculation and results that clearly shows the finite time stable results by this strategy are far better than bounded results by applying previous strategies(Sun, Zhao, Sun, Xia, & Sun, 2020).

**Event-triggered Output Feedback Control for a Class of Uncertain Nonlinear Systems**

In this research, the researcher has discussed the event triggered output feedback for said uncertain non linear systems. Two new event triggered strategies namely fixed threshold strategy and relative threshold strategy are proposed which do not require input-to-state stable which was widely used before. Apart from these new strategies, a new encoding-decoding technique for event triggered control signals is introduced in which only one bit signal either 1 or 0 is transmitted whenever triggering event is violated which further reduces the load on communication channel and consumes less bandwidth. Also the results are verified by Lyapunov analysis. Simulation result shows that the output signal can be regulated around zero while all other signals get bounded. Future problems are also mentioned which needs further study to eliminate remaining errors(H. Li, Liu, & Huang, 2021).

**Extended Fuzzy Adaptive Event-Triggered Compensation Control for Uncertain Nonlinear Systems With Input Hysteresis**

In this research, the researcher has discussed the problem of adaptive event triggered control in systems having input hysteresis. The traditional techniques do not consider the input hysteresis in above systems and do not save communication resources effectively. To overcome this problem, a new method of extended fuzzy approximation is proposed in which time varying error is approximated. Along with this approximation, an adaptive event triggered compensation control with input hysteresis is designed which can effectively compensate for input hysteresis and save communication resources. At the end, two simulation examples with experiments are given which guarantees the system performance and effectiveness of the proposed system(Chen, Wang, Wang, & Wang, 2020).

**Event-Triggered Adaptive Output Feedback Control for a Class of Uncertain Nonlinear Systems With Actuator Failures**

In this research, problem of event triggered adaptive output feedback is considered in the said systems having actuator failures. An event based output feedback controller is designed by using adaptive back-stepping technique in which Neural networks are introduced. Also event triggering mechanism is designed using time variant threshold strategy which helps in making even triggered output feedback controller for bounding the signals of closed loop system. An example of simulation is given at the end to compare the results of proposed technique with previous known techniques. The result clearly shows the triggering events are reduced and proposed strategy has better tracking effect. It also can handle the coupling due to loss of actuator failures and unknown control direction(Zhang & Yang, 2020).

## Section II

**Event-Triggered Output Synchronization of Heterogeneous Nonlinear Multi-Agents**

The research proposes a solution for the synchronization problem that occurred in a heterogeneous non-linear multi-agent system. The primary technique used for distributed event-based controllers was a two-step synchronization process. The author first outlined the consensus controllers for linear models. The further subsequent triggering events were designed based on intermittent communication. To design controllers for an event-triggered perturbed response, the input-to-state stability property technique was utilized. The technique is based on the assumption that each agent involved in input or actuator disturbance is associated with some input-to-state stability (ISS) property. The technique is suitable for output synchronization problems related to heterogeneous nonlinear multi-agent systems developed for networks with directed topology. The Zeno phenomena can be avoided and monitoring problems can be reduced by using this technique. Moreover, and the model is equally suitable for non-linear and linear systems(Khan, Chen, & Yan, 2019).

**Event-triggered controllers based on the supremum norm of sampling-induced error**

The study provides an event-triggered control scheme for interconnected systems. It was first developed by considering auxiliary input and output pairs associated with closed-loop systems. The auxiliary input and output can be defined as sampling disturbance, and derivative of continuous input function respectively. The input is also described as an integral of auxiliary output in sampling. The auxiliary input and output cycle are completed via integral and system dynamics for input and output respectively. It formed an event-triggered law for stabilized mapping with easy to check Zeno-free conditions. The scheme thus formed utilized to develop a theorem for event-triggered control for non-linear and interconnected systems. The scheme was proved to be applicable even for lower-triangular systems having dynamic uncertainties(Zhu, Chen, Hill, & Du, 2020).

**Event-Based Practical Output Regulation for a Class of Multiagent Nonlinear Systems**

The study focuses on multi-agent nonlinear systems and explores the problems associated with practical and cooperative output regulation issues. The authors did so by applying an event-based output feedback scheme. The primary focus of the study is local measurements based on sampled- data. The heterogeneity of the agents is overcome by stabilization to develop an augmented system. This system is comprised of agents and their relevant internal models of continuous time. The measurement feedback method is applied to solve stabilization problems which assist to establish an authentic event-based protocol. This protocol was applicable for providing Zeno-free conditions with globally practical stability in the system controls(Wang, Sheng, Xu, Chen, & Su, 2018).

**Sampled Measurement Output Feedback Control of Multi-agent Systems with Jointly-Connected Topologies**

The study focuses the multi-agent systems having predetermined measurement output within the sampling. The consensus problem of these systems in linear dynamics is investigated and problems are identified. The agents data transfer occurred in periodic manner and thus need discrete-time control for sampled-data. The networks are assumed to have jointly-connected topologies and switching for a certain time interval. The authors propose a theoretical framework to resolve consensus problem and achieve full state consensus for a particular and small sampling time interval. The framework also assists in managing sampled measurement output feedback control for multiple agent systems. A class of jointly connected conditions is applied within the framework to minimize the prohibitive network supposition. The framework is supposed to manage continuous-time multi-agent systems for output feedback control. This model is equivalent to discrete-time model in data sampling and addressing and can be applied for switching topologies. The agents with second-order dynamics are specifically managed by the framework as discrete-time model served as a base for the explicit development of consensus controllers for these agents. The displayed numerical reenactment confirms the accuracy of these controllers. The consensus for five agents for position feedback control is achieved as proposed in this research. The observation portrayed in Fig. 1 is steady with the theoretical computation of (X. Chen, Chen, & Mei, 2016). The profiles of positions and speeds (in x-axis point) of the five agents with discretionarily chose starting states are displayed in Figure 1. The profiles in y-axis are comparable and subsequently overlooked.

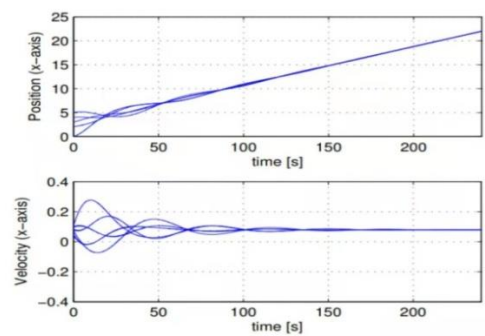

Fig. 1. Profiles of the agent positions (top) and velocities (bottom) in x-axis

The consensus condition of of continuous-time system is comparable to the discrete-time system with asymptotic stability for a sampled position feedback. Asymptotic stability of the discrete-time system can be achieved by an explicitly planned controller when the network is outfitted with switching but jointly connected topologies.

**Performance Analysis of Nonlinear Sampled-Data Emulated Controllers**

The goal is to foster another methodology for explicit execution investigation of nonlinear sampled-data controllers. This new methodology raised from the partner advancement for linear systems. Over the previous years, there is considerable improvement in controlling of sample data for linear systems in, e.g., (Kao & Lincoln, 2004; Mirkin, 2007). Delaying input is used to manage time dependent delay and clarification of samplings. An assortment of continuous-time controllers' dependent on the high gain mastery approach and the back stepping strategy used for diverse range of interconnected systems. A suitable process is emulation used for processing of sampled-data control systems. It is planned to use a continuous time controller in a continuous-time frame system and discretize the controller for digital execution. They stretch out the input delay way to deal with nonlinear sampled data systems to accomplish the objective by permitting time-varying delay and arbitrary sampling patterns. It is difficult to produce a digital nonlinear robust carrier with continuous time control because it required high continuous gains which is related to sampling bandwidth. The methodology permits uncovering the quantitative relationship among region of attraction, controller gain, robustness, and sampling bandwidth for explicit nonlinear systems. We can see from the simulation of the example consider in the research that the SD/D controller functions as a system stabilizer. In any case, it might bomb when a higher gain is utilized. The perception portrayed in Fig. 1 is predictable with the hypothetical estimation of (Zhiyong Chen & Fujioka, 2014).

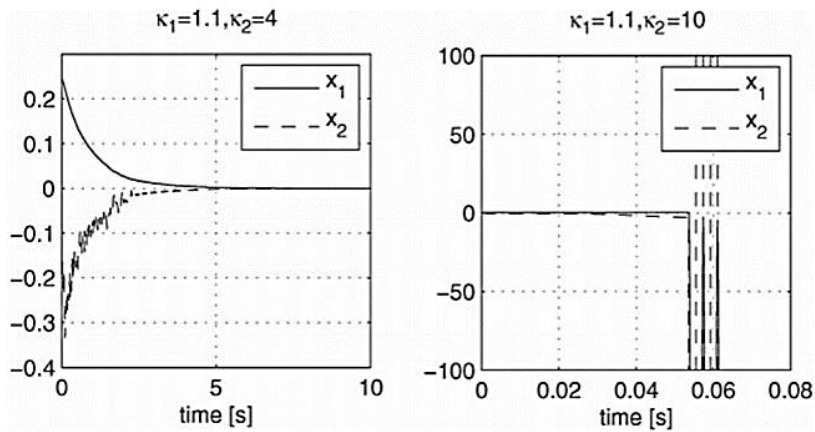

Fig. 2. The designed closed-loop system and its trajectories

The technique has been used on various nonlinear lower triangular systems to produce stability conditions in a sampled-data with delay control situation. The robustness considered inside this structure is concerning obscure system parameters as well as unmodeled delay and nonuniform sampling patterns.

**Robust Sampled-data Output Synchronization of Nonlinear Heterogeneous Multi-Agents**

This examination concentrates on the study of synchronization issues in nonlinear heterogeneous multiple agents with distributed sampled-data controllers. The methodology proposed in (Zhiyong Chen & Fujioka, 2014) is extended for the sampled-data stabilization for perturbed sampled data output regulation problem. It is notable it incorporates an interior model that frames a dynamic compensator which is a procedure for sampled-data emulation of a dynamic controller. Nonetheless, the sampled data emulation approaches just compelling for a static controller. It brings the primary specialized trouble. Likewise, the methodology doesn't oblige the situations with outer perturbation which anyway shows up in the perturbed sampled-data output regulation problem. It is the second specialized trouble. These two challenges will be defeated in another system. The really true exploration is to propose an all the more essentially relevant sampled-data control algorithm for the perturbed output regulation problem and the algorithm further suggests a class of distributed sampled-data controllers for the synchronization problem of nonlinear heterogeneous multi-agents. Although, the synchronization problem of the output transformed into minor perturbed output as an individual agent is incorporated into the referenced models. References tracking can be performed on the perturbed output regulation problems within the sight of considered obscure boundaries dependent on references generated by bound perturbation plagues exo-systems. To solve the perturbed output regulation problem, an inward sample data based dynamic sampled data controller is proposed. An asymptotic bound on the tracking error as a function of the perturbation's greatness and sampling period is determined. In simulation of numerical example considered in this research. Fig. 3 shows the plot of five agents when the time is 1 sec. The output waveform shows that the agents have achieved the required level of synchronization and the output has transformed into a pure sinusoidal wave. Fig. 4 shows that output at a magnified scale of $10^{-5}$ seconds. At this magnification the failure of synchronization and the flops are visible. The understanding is illustrated in Fig. 3&4 is reliable with the hypothetical computation of (X. Chen and Z. Chen).

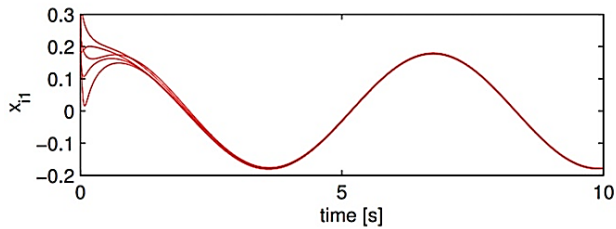 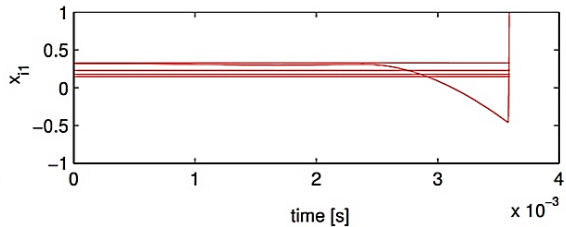

Fig.3. Synchronization profiles of the agents (sampling period 1 ms)    Fig.4. Unstable profiles of the agents (sampling period 5ms).

The procedure has the potential for dealing with time delays which is a fascinating theme for the future exploration.

## 2. Conclusion

The event-triggered control for both linear and non-linear systems has investigated by many researchers. The issues of heterogeneity and Zeno are examined and solutions are being proposed for the said problems. The solutions based on Global Finite-Time Control and Feedback Control Output are proposed for uncertain non-linear systems. Moreover, the issues of output synchronization and problems related to multi-agent systems are discussed to find appropriate models to mitigate these deficiencies. The input-to-state stability (ISS) property technique is proposed by various researchers to manage heterogeneity and synchronization in systems. The latest development in this research area has enhanced the reliability and adaptability of event-triggered control systems as almost all the associated problems and drawbacks have been investigated. There are a lot of proposed solutions for each particular problem in the non-linear event-triggered control system that can be utilized individually or in combination with other suitable techniques to get desired outcomes.

**Acknowledgements**

It is declared that this research is self-motivated and no financial help is taken from any organization for which the rights needs to be granted.